# WHAT MAKES A SYSTEM COMPLEX?
# AN APPROACH TO SELF ORGANIZATION AND EMERGENCE

Michel Cotsaftis
LACSC/ECE mcot@ece.fr

> "Men in their arrogance claim to understand the nature of Creation, and devise elaborate theories to describe its behaviour. But always they discover in the end that God is more clever than they thought
>
> Sister Miriam Godwinson

The fast changing reality in technical and natural domains perceived by always more accurate observations has drawn the attention on a new and very broad class of systems mainly characterized by specific behaviour which has been entered under the common wording "complexity". Based on elementary system graph representation with components as nodes and interactions as vertices, it is shown that systems belong to only three states : simple, complicated, and complex, the main properties of which are discussed. The first two states have been studied at length over past centuries, and the last one finds its origin in the elementary fact that when system performance is pushed up, there exists a threshold above which interaction between components overtake outside interaction. At the same time, system self-organizes and filters corresponding outer action, making it more robust to outer effect, with emergence of a new behaviour which was not predictable from only components study. Examples in Physics and Biology are given, and three main classes of "complexity" behaviour are distinguished corresponding to different levels of difficulty to handle the problem of their dynamics. The great interest of using complex state properties in man-made systems is stressed and important issues are discussed. They mainly concentrate on the difficult balance to be established between the relative system isolation when becoming complex and the delegation of corresponding new capability from (outside) operator. This implies giving the system some "intelligence" in an adequate frame between the new augmented system state and supervising operator, with consequences on the canonical system triplet {effector-sensor-controller} which has to be reorganized in this new setting. Moreover, it is observed that entering complexity state opens the possibility for the function to feedback onto the structure, ie to mimic at technical level the invention of Nature over Her very long history.

## 1. Introduction

After a very long period of observations, it became very slowly clear to first Human observers since prehistoric age that the phenomena they were distinguishing were due to a specific order in Nature and not to the capricious will of mysterious Gods. Since then, they have been over the centuries patiently organizing the most visible ones, and, proposing hypotheses to find this order, they finally end up on laws to represent the simplest phenomena first which have been verified in the framework of actual observations. These elementary phenomena were mainly concerning "simple" systems which can reasonably be isolated in their dynamics and their observation. In the mean time, the advance in technology and observation accuracy was driving the attention on more complicated systems with always larger number of elements, which in the 17$^{th}$ century have been shown by R. Descartes to be reducible in their study to the former class of simple ones[1]. Finally, the result today of this long quest was an "adapted" representation of the Universe with a broad "classical" part for "human size" phenomena, corrected by "quantum" effects at very small atomic level, and by "relativistic" ones at very large galactic level. This basic "1-dimensional" picture emerging from millennium long effort of human kind, and leaving fundamental questions only at infinitely large universe and infinitely small super elementary particle levels, is however not sufficient to represent correctly enough the present situation where systems exist over an extremely large parametric domain now easily explored with the recent development of always more performing technology over the last decades.

First observed from time to time in different domains, there is now evidence of existence of new huge class of systems, natural as well as artificial in all scientific and technical human activities, which have reached their own status by the corner of the millennium under the name of "*complex*" systems. They are in some sense opening a "2-dimensional" more global picture of the universe by being located along a direction "perpendicular" to previous 1-dimension line going from infinitely small to infinitely large size, see Fig.1. There is today a strong questioning about their origin and their formation[2]. Some have even suggested that previous "mechanistic" approach to the universe by researching "ultimate" fully unifying law[3] is just mistaking their very and irreducible structure and that a new "revolution" is necessary to grasp all aspects of known universe today[4]. To address these questions, a very pedestrian approach is proposed here, based on elementary source-sink model applied to the graph representing the aggregate of possible components of a system. Then it evidently appears that system structure can be divided into three different groups, simple, complicated and complex, corresponding to specific properties. The first two groups are usual ones nicely approachable by the methods of scientific reductionism[5]. The third group, by its very global nature, cannot be just reduced to the effect of its components[6] and requires some adjustment for being correctly handled, because now the key point is the way the system behaves under (or against) the action of its environment. It will appear that the mechanistic notion of individual "trajectory" is loosing its meaning and should be replaced by more general "manifold" entity corresponding to accessible "invariants" under environment action. System self-organization and emergence properties are deduced and discussed. Different grades of complexity are fixed depending on specific system properties which relate to well known classes of observed phenomena. Advantages for application of complex structure to man made systems are stressed.

## 2. System State Analysis

Let consider first a system with a finite number $N$ of identified and distinguishable components which can be represented by a graph with $N$ nodes $N_i$. On this graph, three types of vertices can be drawn in between the $N$ components $i$ and outside sources $e$ whenever an exchange exists between them. They correspond respectively to free flight state for vertex $V_{ii}$, to driven state from outer source for vertex $V_{ie}$, and to interactive state with other system components for vertex $V_{ij}$, see Fig.2. System dynamics are usually resulting from a combination of previous three different exchanges to which, on a very general setting, there can be associated three characteristic fluxes for each system component. Their specific nature (power, information, chemical,..) is depending on the problem at hand, but is here unnecessary as long as they unambiguously characterize system components status. The first flux corresponds to the ''free'' dynamics of i-th component $p_{ii}$ along vertex $V_{ii}$, the second one $p_{i,e}$ to the transfer flux between outer source and system i-th component along vertex $V_{ie}$, and the last one to inter components effects $p_{i,int} = \text{Inf}_j\{p_{i,j}\}$, where $p_{i,j}$ is the characteristic and oriented flux exchange between components i and j along vertex $V_{ij}$, see Fig.2. Depending on the various possible ordering between these three fluxes, the system behaves in different ways. When (A) $p_{ii} \gg p_{i,int}, p_{i,e}$ the i-th component is weakly coupled to other components. Its dynamics are mainly the ones of an isolated element slightly perturbed by the other actions, and the system reduces to a set of almost independent one-component sub-systems. In this sense it can be considered as a **simple** one. When (B) $p_{i,e} \gg p_{i,int}, p_{ii}$ the i-th component is mainly depending on outside sources. The action of the other components creates a (weak) coupling between the components, but the system can nevertheless be decoupled, at least locally, into a set of sub-systems which can be acted upon and controlled by as many exterior sources as there are components in the system because they can still be identified. In this sense the system can be termed as **complicated** from etymologic sense (from Latin *cum pliare* : piled up with). Finally when (C) $p_{i,int} \gg p_{ie}, p_{ii}$ the i-th component is very strongly coupled to other ones and its dynamics are now mainly determined by components interaction satisfying the inequality. If apparently there does not seem to exist a dramatic difference with the previous case as concerns their dynamics, there is however a fundamental one as concerns the effect of outside action. In previous case input from the source can be tracked to the concerned component so this degree of freedom can be completely controlled from outside, whereas in present one the nature of interactions dominates and shields this tracking. As a result, control action can only be a ''global'' one from other system components satisfying condition for second case (B) and system dynamics are now also driven by internal action. Consequently a totally new situation occurs for all system components passing in third state in the sense that their control cannot as in second case be fixed by only outer source action. Effectively, it is no longer possible as in second state to determine completely all system components dynamics from outside because of stronger interaction effect which now dominates the dynamics of concerned components. In a sense a *self-organization* has been taking place inside the system which leads to an internal control replacing the classical one from outside. So it is no longer possible to continue to manipulate inputs with as many degrees of freedom as the system initially has because of the mismatch which would result from the conflict with the internal control due to components interaction satisfying condition for the third case. External system control dimension is thus *reduced*. A system in this case will be termed as **complex** in agreement with etymology (from Latin *cum plexus* : tied up with). According to this definition, a very elementary test for determining if a system is passing to complex state is thus to verify that its control requires the manipulation of less degrees of freedom than the system has initially. There exists an immense literature about complexity, its definition and its properties covering an extremely wide range of domains from Philosophy to Technology[7], especially in recent years where its role has been "discovered" in many different fields such as networks now playing a crucial role with the ascent of Information Technologies[8]. Here emphasis is more modestly put only on more restricted complex state compared to complicated state as concerns action from outside environment (ie from control point of view) onto the system. For a system **S** with a finite number $N$ of identified components, one can then define its *index of complexity* $\mathcal{C}_S = 0$ if it is only in complicated state, and $\mathcal{C}_S = n/N$ if $n < N$ components cluster in complex state by satisfying inequality (C) and become insensitive to outer action. $\mathcal{C}_S$ measures the grasp on the system from outside action/control and the limit $\mathcal{C}_S = 1$ corresponds to totally autarchic system, the most complex possible structure with this number of components. This may look paradoxical as the definition of a simple system is precisely that it can be isolated. In fact this apparently contradictory statement is resulting from the very nature of internal interactions effect which reduces the number of invariants on which system trajectory takes place. An elementary example is given by the particles of a neutral gas for which their initial $6N$ positions and velocities (the mechanical invariants of motion) are reduced to the only energy (or temperature), justifying thermodynamic representation. In other words, when represented from outside, a system is the less complicated as it is more complex inside. Differently said, a complicated system remains complicated whether observed from inside or from outside, whereas when becoming complex it is less complicated when observed from outside, on top of the fact that being depending on a restricted number of parameters, it is more insensitive to outer action. However, when observing from outside a system with apparent $\mathcal{C}_S = n/N$, if $N$ is not fixed when the system is not well known, there is no one-to-one relation in general between apparent outer behaviour of the system and its initial internal structure. It is only possible to check relative variation of $\mathcal{C}_S$ when crossing condition (C) for some components. On a general setting, it is very easy from its definition to envision complex state as a powerful way set up by Nature to generate more sophisticated entities able to gain at the same time more independence with respect to their environment, or, in other terms, to become more robust to its variations and yet, to appear much less complicated outside, so that they can in turn link more easily to similar entities and generate a chain of nested clusters. This seems to be an underlying trend in evolution on Earth which has been allowing the elaboration of living beings, the up-most advanced form of complex systems[9].

Summarizing, exactly like matter does appear in three states : solids, liquids and gas, depending on conditions, systems are exhibiting three states : simple, complicated and complex. A very simple way to locate a system with respect to them is to plot for each component i the three values $\{p_{ii}, p_{i,e}, p_{i,int}\}$ in a three dimensional space where each direction represents a single state. Each component is then represented by a point $C_i$ and the system by a cluster of N points, see Fig.3. Depending on where they are located in this space, it is evident to figure out the status of the system by inequalities (A),(B) and (C). Moreover, it is easy to check what will be the consequence of parameter variation, especially when crossing inequality (C), a point extremely important for control of man-made systems as seen later. In this picture, components for which inequality $p_{i,int} \gg p_{ie}, p_{ii}$ holds are in fact an internal sub-cluster which cannot be further split from outside observation and action. As indicated above, the first two states have been observed and studied along past centuries. The fact is that with modern and very detailed observation diagnosis, complex state has been now very often observed in natural systems, and becomes the most common one in a broad range of phenomena. For artificial man-made systems, even if it were sometimes crossed in the past, it is only with recent progress of advanced technology that, when pushing systems performance for higher efficiency, the threshold for overtaking of internal interactions effect corresponding to third case is also very routinely over passed now. So complex behaviour is seen in many situations, and becomes an object of study in itself, with considerable consequences in technical and industrial applications it implies in conjunction with the corresponding demand of more delegation of decisional nature to the systems for better efficiency.

From previous and relatively rough source-sink representation always applicable in a first instance, it is already possible to stress a deep difference existing between the first two states and last complex one. In first two states, the possibility exists to split the system into as many independent one-component systems in a first approximation, whereas this is impossible in third state where all interacting components have to be taken as a whole. In mathematical terms the consequence is that usual approximation methods developed for the first two states do not straightforwardly apply and have to be revised in order to handle the global aspect of the coordinated response of components in complex state. This difficulty is at the origin of today important computerized research undertaken on the problem.

## 3. Emergence

When going to more specific situations, other elements are also entering the description and it is interesting from elementary global description above to recover various situations which have been observed and analytically studied for different parameters value. First it is likely that in general the system is not always in a ''pure'' state and often exhibits a mixed structure where some components are in one state and others in another one. An important example is the case of natural inhomogeneous but continuous natural systems such as fluids with non zero gradients in a domain. In these systems, fluctuations are universally observed the source of which is the free energy available in between this stationary equilibrium and complete (homogeneous) thermodynamic one. In present case, the free energy is coming from the space gradients related to medium non homogeneity. They have in general a very large range of fluctuations (roughly because the system has a very large number of components) which can be split into two groups depending on their wavelength compared to system characteristic gradient length. Typically, the fluctuations with large wavelength excited in the medium are in complicated state and, because they are sensitive to boundary conditions, can be observed and possibly acted upon as such from outside. Fluctuations with small wavelength on the other hand are generally in complex state due to parameter values. So under their strong interactions, and because they are much less sensitive to boundary conditions, they are loosing their phase and globally excite a leak out of the fluid (usually called a *transport*) manifested by an out-flux expressing the non equilibrium situation of the system, and which counteracts the input flux responsible of medium non homogeneity. The determination of these transports is a very important element for qualifying the behaviour of the system and is an active research problem studied worldwide. This feature is observed for all natural systems when they enter the *dissipative* branch.

A natural system is called dissipative[10] when it is exchanging (particles and energy) fluxes with outside environment. Evidently the channels by which internal energy sources are related to these fluxes are playing a privileged role because they regulate the energy ultimately available for the system and finally determine its *self-organized* state[11]. Dissipative systems more generally only exist to the expense of these fluxes, and they evolve with parameter change – such as the power input – along a set of neighbouring states determined by branching due to bifurcations where internal structure changes in compatibility with boundary conditions and by following the principle of largest stability. So the picture of such systems is a transport system governing flux exchanges guided by the bifurcation system which, as a pilot, fixes the structure along which these exchanges are taking place. Finding the branching pattern thus entirely defines the possible states of the system and determines the fluctuation spectrum. Branching is found as nontrivial solutions of variation equations deduced from general system dynamics equations. Despite the physical origin of phenomena is here well identified, the analysis is still in progress in a very large number of situations, such as in thermonuclear plasma to determine correct parameter sizing for realizing "burning" conditions[12], and in neutral fluids where turbulence is not fully explained yet[13], whereas the problem has been cleared up for fluctuations in deformable solid bodies[14]. In most cases the difficulty stems from finding an adequate representation of the global effect of small wavelength fluctuations, especially when due to parameter values, they feedback on longer time scale large wavelength oscillations and modify their dynamics. In this case, there is a significant change of initial system dynamics due to passage in complex state of an internal part of the system generated by branching phenomenon. Moreover, it is very likely that modification of system dynamics is the more important as the non homogeneity zone is the less ''transparent'' between the two domains above and below along power flow. This has been extended to the extreme with living cell systems which have been able to completely encapsulate within a filtering membrane (ie a steep gradient) a space domain where very specific "memory" DNA molecules are fixing the dynamics of inside system they control, with corresponding exchange across the membrane. It is easy to understand that conjunction of a gate and of adequate information is the most efficient way to deeply modify locally the effect of regular physics laws, as exemplified by the counter streaming motion of a piece of wood in a river with one stage locks when regulating their opening and closing.

However, dissipative behaviour does not exhaust all possible situations and many other ones do not follow this pattern. Natural systems so far are exhibiting components with relatively elementary features (charge, mass, geometrical structure, chemical activity, wavelength, frequency..), but there are also cases where complex state occurs in systems with more sophisticated components, usually called "*agents*". Examples are herds of animals, insect colonies, living cell behaviour in organs and organisms, and population activity in an economy. In all cases, when observed from outside the systems are exhibiting relatively well defined behaviours but a very important element missing in previous analysis is the influence of the *goal* the systems are seemingly aiming at. Very often the components of these systems are searching through a collective action the satisfaction of properties they cannot reach alone, and to represent this situation the specific word "*emergence*" has been coined[15]. The point is that it is now possible to return back to previous case and in a unified picture to envision the laws of Physics themselves as emerging phenomena. For instance for an ensemble of neutral particles with hard ball interactions, and beyond the threshold of rarefied gas, (ie when the Knudsen number Kn = $\lambda/L$ is decreasing to 0 from the value 1, with $\lambda$ the particle mean free path and $L$ a characteristic length scale), the particle system is suddenly passing from a complicated to a complex state (due to the huge value of Avogadro number). Consecutive to overtaking of collective interaction by collisions (expressed by decrease of Knudsen number), it could be said that there is emergence of a pressure and a temperature, which, from a point of view outside the gas, summarizes perfectly well the representative parameters (the invariants) describing it at this global level and amply justifies the thermodynamic representation in this case, see next §. The elementary reason is that here the fast particle dynamics described by Boltzmann equation relaxes within extremely short collision time to its equilibrium Maxwell-Boltzmann-Gibbs distribution only depending on temperature (and the potential of applied forces if any), which justifies usual stochastic methods[16]. Of course it is tempting at this point to "forget" the underlying mechanical base and to decide that perfect gas law, if any, is the emergence directly resulting from billiard ball "properties" of interacting (agents) molecules. The difficulty in this case is that application of "ergodic" hypothesis[17], even if it can mathematically be invented from scratch, rests here heavily upon thorough analysis of molecule ensemble dynamics, ie on mechanical laws. Similar problem does exist at atom level itself where, after baryons are assembled from primitive quark particles below some threshold energy, protons and neutrons assemble in turn themselves below another lower energy threshold into ions with only mass and charge parameters, able to combine finally with electrons to create atoms. Speaking of "emergence of atoms" from underlying components does not add any more information to classical approach. So in an opposite direction, a more conservative and probably more appropriate way to account for the new situation is to just state that when becoming complex, systems are exhibiting emergence of self-organization out of which it has been verified that their new behaviour in complex state is always a consequence, without going into the details of this behaviour which is depending on specific system parameters. In fact it is quite elementary from previous source-sink model to understand that a key point is in the accuracy of modelling the components in complex state, as long as the resulting "invariants" which will grasp all system information for interaction with environment are directly depending on this modelling. This has been at the origin of a computer "blind" search where the agents are given properties and "emerging" behaviour is obtained in a bottom-up approach, sometimes in surprising compatibility with experimental observations[18], in parallel to theoretical analysis[19]. Of course following classical science reductionism, there remains to justify the choice of agent properties, and to show that they are in one-to-one relation with observed behaviour. As this does not seem always possible, a different holistic approach[20] has been proposed where the research of ultimate underlying elements from which everything is constructed by as many layered shells on top of one another, is considered as unnecessary, to the price of loosing prediction power.

Finally an extremely important remark is that the logical chain :

{stimuli/parameter change} → {higher interactions between some system components} → {passage to complex state} → {system self organization} → {emergence of new behaviour} discussed here is nothing but the sequence leading to the final step of system evolution toward more independence, and which is the feedback of "*function*" onto "*structure*", a specific property of living organisms explaining their remarkable survival capability by structure modification.

## 4. Mathematical Analysis of Complex Systems

The few previous examples from common sense observation illustrate the elements which have been described above and which are providing a general base for complex system paradigm. From atomic nucleus to galaxy natural systems are seen to be constituted by aggregates of identifiable components (which, as already stressed, can be themselves, at each observation level, aggregates of smaller components) with well defined properties. These aggregates have been said to exhibit a complex behaviour when interaction between the components –or some of them– is overtaking their interactions with exterior environment. Similarly living beings are exhibiting the same behaviour, as observed with gregarious species, and in artificial man-made systems the same phenomenon is occurring when the overtaking conditions are satisfied. This is for instance the case for high enough performance level systems because the components are then tightly packed, as for high torque compact electrical actuators. Despite an extremely large variety of possible situations there are few base interactive processes leading to complex self-organization. Three main types will be discussed: the reducible case, the instantaneous differential case, and the integro-differential case with past neighbourhood memory.

4A – Reducible case :

It is first intuitively expectable that the approach to full problem gets easier when time-space scales are very different for components going to complex state as compared to the ones of other system components. This is mostly the situation in dissipative systems where characteristic frequency and wavelength of the spectrum of complex system part are respectively much larger and much smaller than the ones of the other (non complex) part of the system. The main reason is as explained above that components in complex state are produced in the initial system by the effect of bifurcation determined from variation equations of the system and are corresponding to high mode numbers (and have small enough wavelength). For such modes, their saturation amplitude is fixed by their nonlinear interaction which overtakes the level they would have reached under the only effect of the source and, as well known, exhibit globally a fluctuation spectrum usually very difficult to determine analytically. Let $\tau_s$ and $\tau_{cp}$ be respectively the characteristic response time of the non complex system part and of complex modes part with $\tau_{cp} = \varepsilon^n \tau_s$. The simplest situation n > 1 corresponds to the case where the interactions are strong and frequent enough to kill corresponding modes dynamics. In this case and because there are usually many modes with small amplitude, complex system part dynamics can be reduced to ''thermodynamic'' approximation with a random distribution of fluctuation spectrum. Then the initial system of equations describing the full system splits into a restricted set of identified modes perturbed by a random term. This unifying and powerful model is quite successful in science and engineering[21] but it does not cover all situations. A more difficult and yet common case is occurring when n = 1, for which small O($\varepsilon$) modes dynamics have long enough coherence characteristic time comparable to $\tau_s$. To account for this coherence, asymptotic expansion of small modes dynamics from complex part dynamic equations is required up to first order to be injected in non complex system part dynamics, because $\varepsilon$ order term acting coherently over a time $\varepsilon^{-1}$ produces a non negligible finite order term. The system again reduces to its non complex part but now modified by coherent effects of complex system part dynamics, see Fig.4. The unsolved situations in [12,13] are belonging to this case. However it will nevertheless be termed as *reducible* because full system dynamics can be projected onto non complex system subpart. It clearly illustrates the fact that, given a system with a fixed number of degrees of freedom and associated initial and boundary conditions, this number should increase at bifurcation crossing so other conditions would be required to determine the solution. In fact this is not the case here as new complex modes dynamics can be solved by asymptotic procedure and re-injected into non complex system part which obviously keeps the same number of degrees of freedom. The only constraint is on initial complex modes amplitude which, due to their smallness, obey central limit theorem and are randomly distributed. So the difference with previous case n > 1 is that only initial amplitudes values are random now and follow afterward their own dynamics whereas they were previously random for all time. In both cases the number of final degrees of freedom remains the same, so despite their number has apparently increased at bifurcation, in reality the system is still of the same dimension. The extremely important consequence is that for controlled man-made systems, when their dynamics cross a bifurcation point, the same number of controllers is still required but they should be adapted to explore a new larger function space. The main effect of the bifurcation has been to augment system response to a wider function space, and the motion takes place on a *manifold* of this new space so that at the end the effective number of degrees of freedom is remaining the same. As stressed earlier, this behaviour is the manifestation of system *self–organization*. The double mechanical-thermodynamic feature is extremely general and characterizes complex reducible systems where complex part is added by bifurcation crossing, see Fig.5. Again control of such systems (whether natural or man-made) is not classical because for n ≥ 1 the new modes created after bifurcation crossing, though sometimes observable, are not accessible from outside. Because of their definite effect on system dynamics, a new more balanced approach respecting internal system action has to be worked out, which, very generally, is aimed at replacing inability of control action by insensitivity to variation for inaccessible complex modes. Though apparently loosing some hand on such systems, it has been surprisingly possible along this line to find explicit conditions in terms of system parameters expressing somewhat contradictory high preciseness (by asymptotic stability condition) and strong robustness (against unknown system and environment parts)[22]. In this way system dynamics are finally controlled and asymptotic stability can be demonstrated, but in general the price to pay is a not necessarily decreasing exponential asymptotic type.

4B – Instantaneous Differential Case :

The reducible case mainly deals with systems in partly or totally complex state, as with bifurcation crossing by some varying parameter, and with small parameter ordering between time and space scales, so it is possible to proceed to asymptotic expansion. More generally, a system may be in complete complex state, examples of which are atomic nucleus, herd of animals, and galaxies. Despite their very different space sizes, the systems exhibit always the same base characteristic feature to finally depend on an extremely restricted number of parameters as compared to the aggregate of their initial components. Searching the way to extract directly the remaining ''control'' parameters of such systems from their dynamics is a fundamental issue which today motivates a huge research effort worldwide, especially in relation with information networking. Extensive analytical and numerical study has been developed for differential systems of generic form

$$\frac{dX}{dt} = A(t,X,v)X + \lambda.F(t,X,u) + \mu.S(t,X) \tag{1}$$

where $X = \text{col}(X_1, X_2,.. X_n)$ is system state space, $\lambda, \mu$ are n-vector coupling parameters, and $A(.,.,.)$, $F(.,.,.)$, $S(.,.)$ are three specific $q_1 \times q_2$–matrix terms ($q_1, q_2 \leq n$) corresponding to isolated free flight, nonlinear internal interactions and source terms respectively (the linear and source terms in the right hand side of eqn(1) are here split apart to indicate their respective role). The other variables $v=v(X,t)$ and $u(X,t)$ in $A$ and $F$ function account more generally for the possibility to feedback evolution of $X(t)$ onto their own dynamics as it often occurs in systems when splitting parameters into given and manipulated control ones. For fixed $u(.,.)$, $v(.,.)$, depending on the values of $\lambda, \mu$ components the system will be in simple, complicated or complex state described in §1. When increasing the components of $\lambda$ the system runs into complex state, and it has been repeatedly observed, especially on systems close to Hamiltonian ones[23], that system representative point in n–dimensional state space follows a more and more chaotic trajectory when crossing bifurcation values and at the end fills up a complete domain[24]. Of course sensitivity is largest when the system exhibits resonances, ie is close to conservative, and adapted mathematical expansion methods have to be worked out[25]. Because systems are basically non integrable[26], this is a direct evidence of increasing effect of internal interactions which reduce system dynamics to stay on attractor manifold of degree p < n, so that system dynamics are now layered on this manifold. This also expresses the fact that trajectories on the remaining n–p dimension space are becoming totally indistinguishable (from outside) when taken care of by internal interactions of n–p components going to complex state. So system trajectories reorganize here in equivalence classes which cannot be further split, a dual way to express the fact that there exists an invariant manifold on which system trajectories are lying. It is easy to understand that continuing to control these components by regular previous control[27] worked out for complicated state and specially designed for tracking a prescribed trajectory, is no longer possible and that a new approach is required which carefully respects internal system action due to complex state self-organization. More global methods of functional analysis[28] related to function space embedding in adapted function spaces[29] by fixed point theorem[30] are now in order as shown for reducible case[31], because they are providing the correct framework to grasp the new structure of system trajectory which cannot be fully tracked as before. Basically the method is again to counteract impreciseness in an element by robustness to its variation, a method very largely followed by living organisms.

4C – Integro-Differential Case with Past Neighbouring Memory :

Another important dimension relates to the properties system components are given, and a very influential one is the range of inter-component interaction, because this determines completely the build up of system clustering when becoming complex. Obviously long range interactions are leading to more intricate response with more difficult analysis. Examples are stars in a galaxy, electromagnetic interactions between ions and electrons in a plasma, animals in a herd and social behaviour of human population in economic trading such as stock market with internet link. In all cases a new element is coming from the size of the neighbouring domain each system component is sensitive to, and implies a time extension to past neighbouring components trajectories, see Fig.6. So the resulting complex behaviour is more generally determined by interactive component effects over a past time interval and weighted according to their importance. In this case systems are obeying for $t \geq t_0 > 0$ delayed equations of the form

$$\frac{dX}{dt} = F(t, X(t), I(t), u, d) \qquad (2)$$

$$\frac{dI}{dt} = \int \rho(\lambda, t) G\left(t, X(t), \int_{-\infty}^{t} \varphi(t') X(t - h(t', X(t'))) dt', \lambda\right) d\lambda \qquad (3)$$

with $X = \text{col}(X_1, X_2,.. X_n)$, $I = \text{col}(I_1, I_2,.. I_p)$, which clearly indicates the role of the past through intermediate "moments" $I(t)$ gathering a set of elements $G$ with a weight $\rho(\lambda,t)$, each $G$ being itself constructed on the past history of system trajectories $X(t)$ with fading weight $\varphi(t)$, and where $F(.,.,.)$ :

$\mathbf{R}_1^+ \times \mathbf{R}_n \times \mathbf{R}_p \times \mathbf{U} \times \mathbf{D} \to \mathbf{R}_n$ and $G(.,.,.,.) : \mathbf{R}_1^+ \times \mathbf{R}_n \times \mathbf{C}^0(\mathbf{R}_1^-, \mathbf{R}_n) \times \Lambda \to \mathbf{R}_p$ are specific vectors. Their expression generally include control parameters $u(.) \in \mathbf{U}$ which have to be stated explicitly by control law to fix the structure of eqns(2,3) and uncontrolled disturbing effects often represented by a noise $d(.) \in \mathbf{D}$. Here the delay $h(.,X(.)) < t$ can be state-depending as it appears in the analysis of natural (biological) systems where information from the past can be shielded by the state itself. The mixing from the breath of trajectory coverage and from past memory basically creates a "double" complexity in time and space for system components, slightly reduced by averaging effect of acting moments $I(t)$. It indicates that component trajectory is more constrained as it is constantly depending on neighbouring ones, and a possible way is to represent this collective action as an average potential through a distribution function equation of which eqns(2,3) are the characteristics. Apart few existence theorems[32], study of eqns(2,3) is at very preliminary stage today. Like for eqn(1), their numerical study can be undertaken by fixing design and control parameters in $F$ and $G$ such that some trajectory properties are met. Here also a useful companion approach is in application of embedding theorems by fixed point methods, especially when $F$ and $G$ are satisfying generalized Lipschitz conditions, which often provides an embedding property in Sobolev spaces by use of substitution theorems[33]. More precise statement can be obtained when there exist upper bounding estimates in norm of the RHS of eqns(2,3). For example, if the right hand sides of eqns(2,3) satisfy inequalities of generalized Lipschitz type[34]

$$|F|, |G| \leq a(t) + \sum_{i=1,n} \left[ b_i(t) g_i(Z(t)) + c_i(d_i(t)) k_i(Z(d_i(t))) \right] \tag{4}$$

where $Z = \|\text{col}\{X,I\}\|$, $d_i(t) = \text{Inf}_X\{t - h_i(t,X(.))\}$, and the various functions $g_i(.)$, $k_i(.)$, $a_i(.)$ and $b_i(.)$ are positive in their definition intervals, there exists the following bound on the solution

$$Z(t) \leq \mathbf{H}^{-1}\left\{\mathbf{H}_0 + \text{sgn}(\mathbf{H}) \int_{t_0}^{t} \Phi(s) ds \right\} \tag{5}$$

with $H(Z) = \|\text{col}\{1, g_i(Z), k_i(Z)\}\|$, $\Phi(t) = \|\text{col}\{a'(t), b_i(t), c_i(t) d_i'(t)\}\|$, '= d/dt and $\mathbf{H}(v) = \int^v dv/H(v)$, $\mathbf{H}^{-1}(.)$ its inverse function. The solution $X(t)$ is bounded with $Z(t)$ defined over the complete time interval $t \in \{t_0, +\infty\} \subset \mathbf{R}_1^+$ and for all $Z_0 = Z(t_0) \in \mathbf{R}_n^+$ when $\mathbf{H}^{-1}(.)$ is bounded which generally occurs when the function is sub-linear, whereas there is a conditional bound inside the domain limited by $H_0 + \text{sgn}(H) \int^T \Phi(s) ds = 0$ in $\{H_0, T\}$–space, ie in $\{X_0, T\}$–space, when the function is super-linear. When there exists monomial bounds $g_i(x) = \lambda_i x^p$, $k_i(x) = \mu_i x^p$ and $a(t) = 0$, one obtains the explicit bound

$$Z(t) \leq \left[ Z(t_0)^{1-p} - (p-1) \sum_{i=1,n} \left\{ \lambda_i \int_{t_0}^{t} b_i(s) ds + \mu_i M_i \int_{d_i(t_0)}^{d_i(t)} c_i(s) ds \right\} \right]^{\frac{1}{1-p}} \tag{6}$$

with $M_i = \text{Max}\{1/d_i'\}$, showing a limit in $\{t, Z(t_0)\}$–plane for $p>1$ when the term between bracket is 0. Bounds in eqns(5,6) exhibit double advantage of being analytically meaningful and tractable, and to naturally create equivalence classes within which system dynamics are globally the same. This is exactly the expression of system natural robustness resulting from the passage of some components to complex state which can be here easily exploited. An interesting observation from such an approach is again the emphasis on importance of the manifold on which system trajectories are staying, much more than one specific trajectory which has been seen above not to contain enough information for generating an outer action to control it. Another point is the way stability is now perceived in present setting: rather than researching usual specific decay by classical Lyapounov method (for asymptotic stability), it is more generally the belonging of solution manifold to a given function space determined by its properties, see Fig.7. Obviously the bound in eqn(6) exists over the entire time interval only if the term between bracket on the right hand side of eqn(6) is $> 0$, which implies a relation between bounds parameters $\lambda_i$, $\mu_i$ and initial values $Z(t_0)$. So for a given initial ball $\|Z(t_0)\|$, there exists an uncertainty ball $\| [\Delta\lambda_i, \Delta\mu_i] \|$ on system parameters within which the solution belongs to a manifold $\mathcal{M}$ defined by the RHS of eqn(6), the embedding of which in required function space $S$ can be studied in a second step. This opens the way to get at the same time asymptotic stability and robustness, both useful properties for control of complex systems. The approach can be extended to the opposite and more difficult situation where system uncertainty ball is larger than robustness ball[35].

## 5. Applications to man made systems

Evidently previous properties are of up-most importance when applied to man made systems now appearing in industry. Under economic competition, more advanced systems are conceived and worked out which include an always increasing number of heterogeneous components to be operated all together for production of higher value objects. This would imply to keep complete system mastery by efficient control, which becomes the more illusory as system dynamics cover a larger number of elements escaping from one single centralized control structure, especially by becoming complex due to higher value of coupling parameters. In parallel, to reduce system behaviour fragility, it is also interesting to reduce the number of input control parameters by transforming the system into a (partially) complex one by clustering some components into bigger parts. A trivial illustration of this approach is given by the way shepherd dogs are acting on a herd of cattle in the meadows. If there are *n* animals wandering around, they represent a system with *2n* degrees of freedom (*3n* when counting their orientation). Clearly the dog understands that it is hopeless with his only *2* degrees of freedom to control all the animals, so his first action is to gather all of them in a restricted space so that by being close enough they have strong enough interactions transforming their initial complicated system into a complex one with dramatically less degrees of freedom, in fact only two like himself. Then he can control perfectly well the herd as easily observed. The astonishing fact is that dogs are knowing what to do (and they even refuse to do anything with animals unable to go into this complex stage, ie to develop gregarious potential) whereas today engineers are not yet able to proceed in similar way with their own constructions and to get corresponding benefits in terms of global mastery. This is an immense challenge industrial civilisation is facing today justifying if any the needs to study and to create these complex systems. On the representation of system in Fig.3, this would mean to vary adequate parameters to move the representative points along complex axis in order to decide exactly new system status. In any case, internal non controlled dynamics are taken care of by system self organization resulting from passing to complex state, implying that precise trajectory control is now delegated to system. The challenging difficulty is that comply with new structure, some "intelligence" has also to be delegated to the system, leading for the operator to a more supervisory position[36]. In present case, this is contribution to trajectory management by shifting usual (imposed) trajectory control to more elaborated task control [37], a way followed by all living creatures in their daily life to guarantee strong robustness while still keeping accuracy and preciseness. This illustrates if any the limited possibility of expressions from laws of Physics because they are tightly linking *information flux* related to the described action to *power flux* implied in it. Thanks to the discovery of "memory" DNA molecules, Nature has been very early able to "escape" from the constraint of solely following Physics laws strict causality, by inventing sufficiently isolated systems manipulating information flux as well (ie in storing and releasing it according to a timing fixed by survival goal). In fact resulting entropy decrease from created order cannot be maintained for ever due to their unavoidable transport and energy loss because they cannot exist in complete isolation, so Nature has been circumventing resulting isolated system finite life by inventing reproduction permitting species survival instead. In some sense, Human kind today is faced to a similar problem in the research of higher efficiency for industrial systems. After first development of tools, then to machines, and later to efficient control structure, leading to Mechatronics, a new step is now under way to give man made systems more efficiency and autonomy by delegating more "intelligence" to them, see Fig.8. This implies to search an adequate merging of information flux mastery from recent Information Technology development with power flux mastery resulting from classical long term mechanical development[38].
An elementary illustration is provided by the evolution of Robotics to comply with always more severe industrial constraints. It can be shown[14] that robotic systems are completely fixed by the three parameters $(Jr^2)_i$, $\varepsilon_i$ and $(\omega_d \tau_R)_i$ for each link i, with *J* the inertia moment of the link (including actuator), *r* the gearing ratio, $\varepsilon$ the compliance factor and $(\omega_d \tau_R)$ the ''rigidity'' factor with $\omega_d$ the characteristic deformation frequency and $\tau_R$ the characteristic mechanical frequency. First N-link robots were build up so that parameters $Jr^2$, $\varepsilon$ and $(\omega_d \tau_R) \ll 1$ for each link i. The inequalities can be shown in full generality to decouple link interactions and to keep the system rigid, so the system was in simple state. However slowness (speed is multiplied by $r$) could not be compensated by higher actuator rotation speed, and passing to complicated state, use of decoupling method has been allowing direct drive actuation $r = 1$ but with preciseness problems from inadequate control. This has been improved by developing more robust controllers in broader Mechatronics, especially to improve mastery of compliance and vibrations. Recent orientation with higher nominal parameters is toward more autonomous and more intelligent systems with larger decision delegation shifting new robots toward complex state. This new step implies the introduction of "intention" into the system and not to stay as before at simple action level of following prescribed fixed trajectory dictated by classical control. In such scheme intention is represented by the "task" communicated to the system which now determines its own trajectory. Such evolution is following the heavy trend started from beginning of human activity, as described in Fig.8.

## 6. Conclusion

Thorough examination of natural systems in last fifty years has confirmed the existence of systems exhibiting behaviours which do not fit with main stream scientific laws established from patient observations of Nature over past centuries. These laws are based on mechanistic representation of the Universe born from celestial body motion, ie from identifiable simple objects with well defined trajectories. With the ascent of modern technology, new natural systems have been studied and artificial ones have been constructed, both with very intricate structure implying a large number of heterogeneous components in strong interaction. Application of usual laws is often unable to describe their dynamics, because they stay outside the domain of complicated multi component systems only covered by use of reductionism method. The main reason is in the overtaking of component interaction strength which dominates enough over other effects to force the system to close on itself and to manifest an internal self-organization responsible of its new behaviour. Differently said in elementary terms, the new paradigm is that "increasing interactions between components lead to their isolation" as easily verifiable. Such systems are termed as "complex" from etymology, and their main feature is that components in complex state are internally ruled through this self-organization so that at the end they are less depending on environment action. As a consequence, natural complex systems are structurally more robust than complicated ones as evidenced by observation of living organisms, the most complex existing systems. Similarly, artificial complicated systems can be made more robust by transforming them into complex ones when linking adequately and strongly enough some of their components. As always since human origin, the price to pay for this insensitivity is the needs to replace shielded outer action by larger delegation of "intelligence" to the system for its correct internal driving. Analysis of complex systems dynamics shows, for high enough ability of system components (now called "agents"), the possibility of "emergence" of a new behaviour which is not included into the set of initial components behaviour. Though it could be the way of approaching the new global and larger "2-dimensional" picture of the Universe proposed here, to avoid trivial rephrasing of well known classical results, emergence in this context appears to be more justified when restricting its domain to action of agents teaming up for collective goal they cannot individually reach.

The problem of representing complex system structure build up has been discussed, and three typical structures have been singled out. A− The broad class of reducible structure to which natural dissipative systems are belonging, usually formally amenable to be analytically projected back into complicated structure thanks to very large time and space scale difference between components. Application of asymptotic expansion methods allows to directly get important system quantities such as transport coefficients. B− The ordinary differential structure, thoroughly analyzed theoretically and by computer simulation, which shows when increasing component coupling interaction the transformation of system trajectories into finally fully chaotic and indistinguishable ones filling complete domains of state space, manifesting a layered organization of motion into equivalence classes which are the only reachable system parameters from environment. C−The most difficult delayed structure mainly due to long range component interactions, exhibiting a double complexity in space and time together, and grasping the fundamental aspect of component self determination by properly accounting for neighbouring ones within the range of interaction potential. Origin and main properties of class A have been cleared up by Physics study, but transport problem is not completely solved for continuous systems such as Fluids and Plasmas. Class B has been carefully analyzed from Mathematics, especially in the difficult case of resonances for Hamiltonian systems, but still resists to a full picture from analytical reconstitution of trajectory manifold. Class C remains mainly open as its study did not really begin yet. Functional methods approach in adapted function spaces has been shown to be successful for reducible systems in class A, in that both robustness and asymptotic stability are obtained together. Extension to classes B and C will be discussed elsewhere.

As it is based on elementary source−sink description from Physics of interactions between elements of a system (the agents), and between the system and its environment, complexity paradigm discussed here can be tested through a falsifiable protocol. This bottom-up analysis is directly privileging a middle (control engineering type) approach along possible action on the system from environment in between purely (speculative) mathematical and purely (contemplative) physical classical approaches, because it fits immediately with global concepts of auto-organization and adaptive robustness resulting from transformation to complex, and allows direct application of powerful and particularly adapted modern functional analysis methods. Application to artificial man-made systems is particularly appropriate for highly performing ones, as entering complexity is opening the possibility for the "function" to feedback onto the "structure", due to emergence produced by self-organization, a very specific property of the only living systems to date guaranteeing their remarkable survival ability. Finally, possible extension of this approach to "soft" sciences could open on orientation of economic and social sciences research towards more rigorous and potentially predictive models of human conduct.

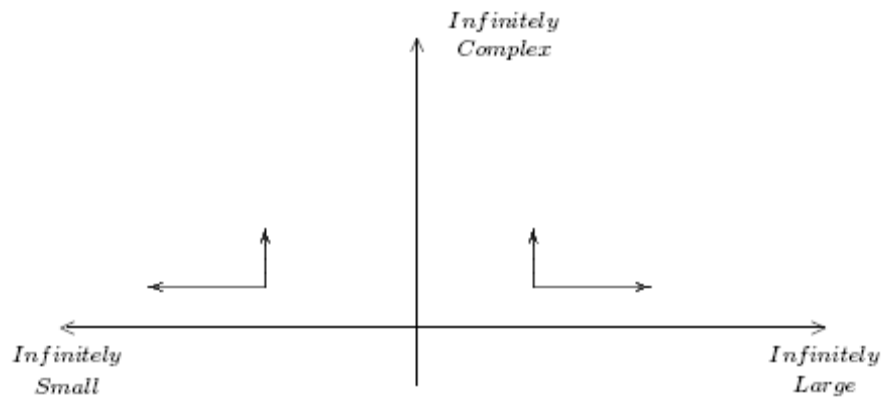

Fig. 1 : New 2-D Representation of Universe with Arrows along Recent Trends

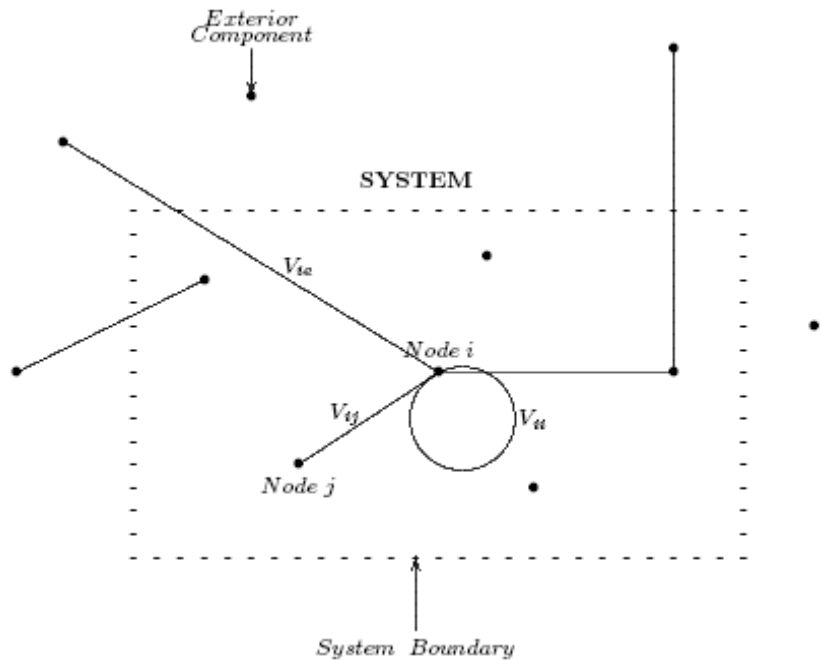

Fig. 2 : Graph Representation of System with
its Three Exclusive Types of Vertices $V_{ii}$, $V_{ie}$ and $V_{ij}$

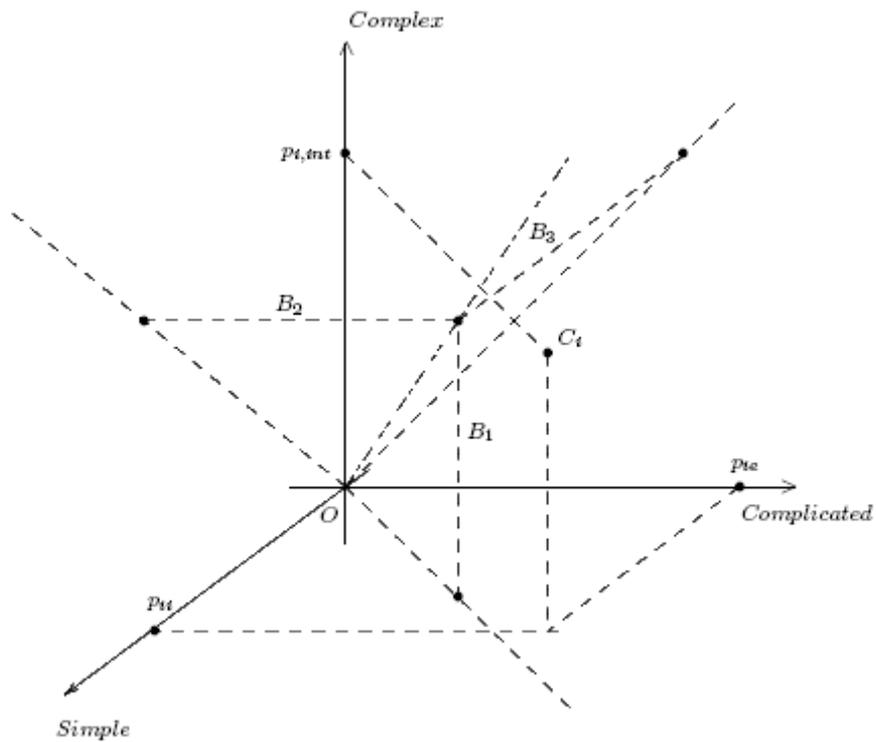

Fig. 3 : System Representation in [Simple-Complicated-Complex] State-Space Domain :
- Inequality (A) is satisfied in Tetrahedral Domain $(0, B_1 B_2)$
- Inequality (B) is satisfied in Tetrahedral Domain $(0, B_1 B_3)$
- Inequality (C) is satisfied in Tetrahedral Domain $(0, B_2 B_3)$
Component $C_i$ is in Complicated State

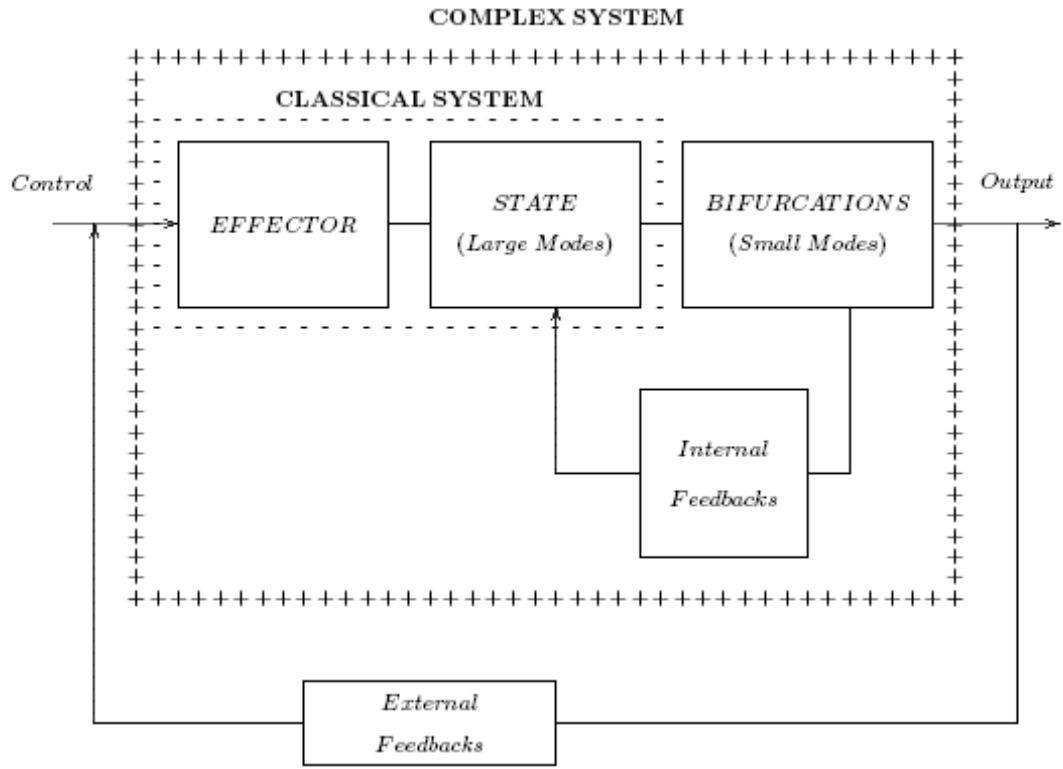

Fig. 4 : Block Structure of Typical Reducible Natural Complex System

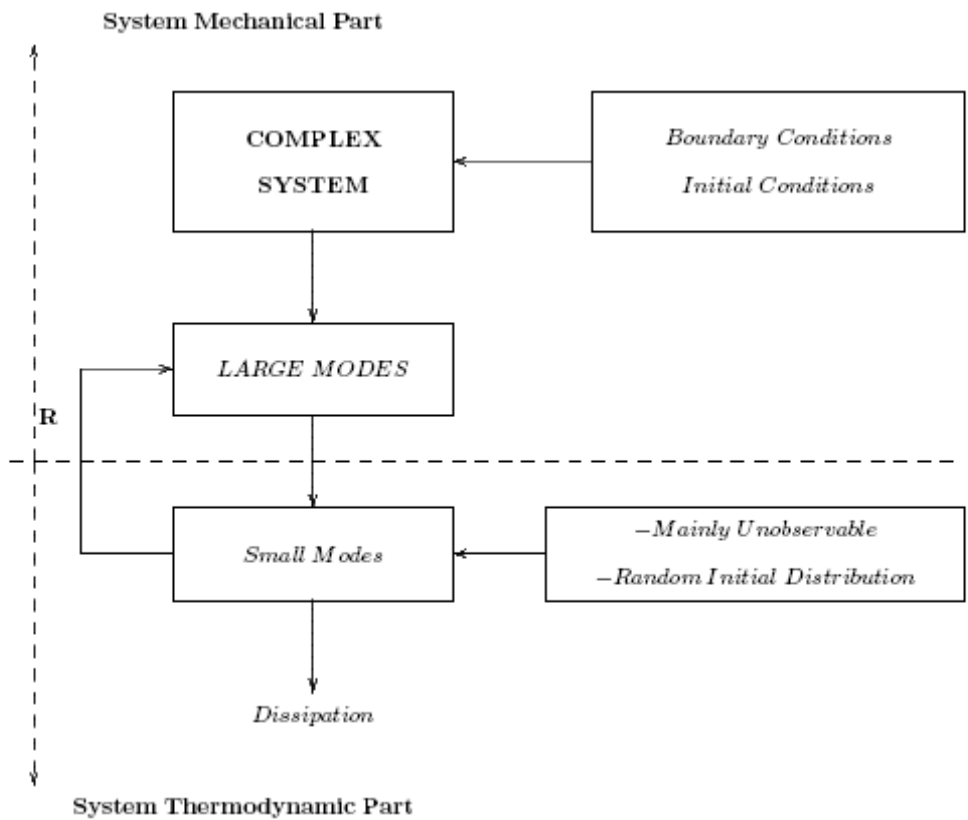

Fig.5 : Sketch of General Split Complex System Block Scheme

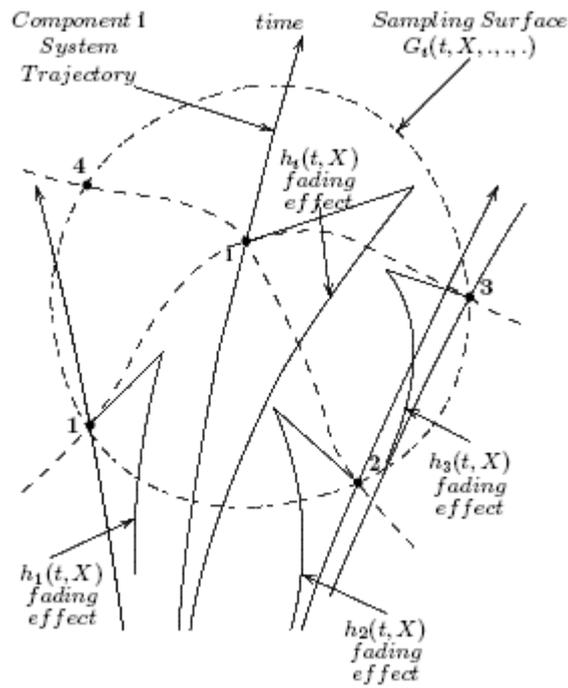

Fig. 6 : Domain Covered in Trajectory Space by Sampling
Function $G_t(\text{— -})$ with Components 1,2,3,4 and
Normalized Past Trajectory Memory Effect Due to Delays $h_j$

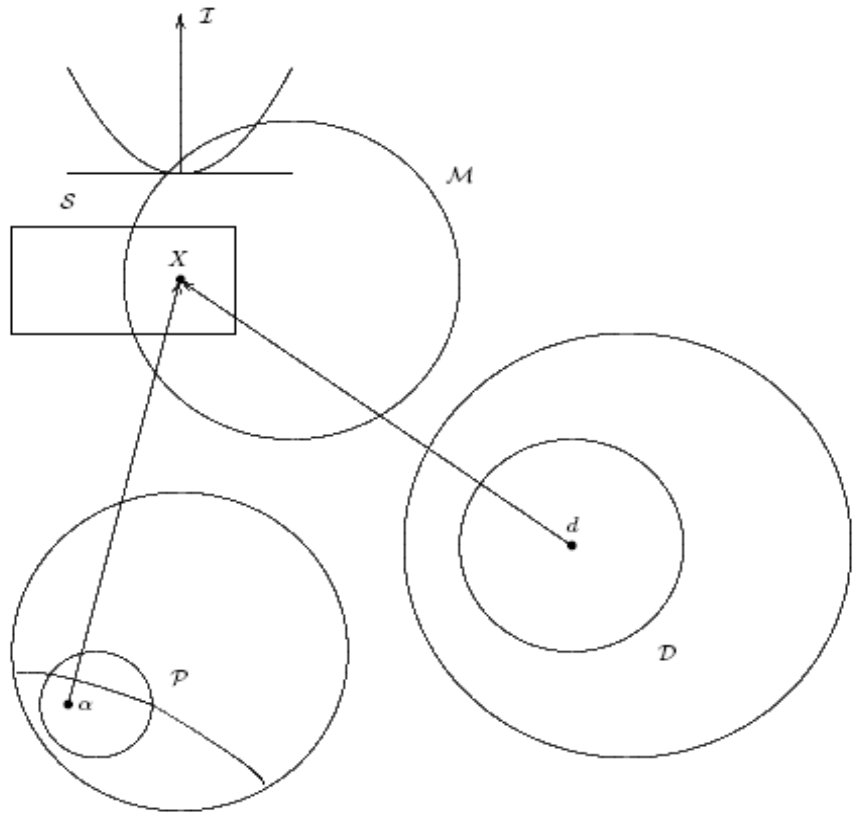

Fig.7 : Embedding of System Solution in Preselected Function
Space $\mathcal{S}$ while Optimizing Trajectory Functional $\mathcal{I}$

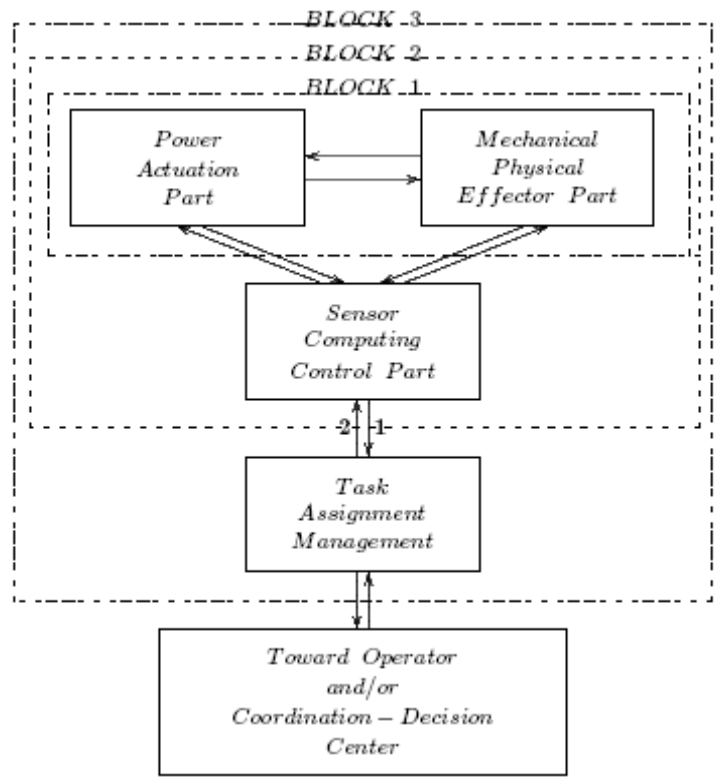

Fig.8 : System Structure Evolution with Main Added
Component Parts. Flux 1 Corresponds to Useful Information Transfer.
Flux 2 Corresponds to Intelligence Delegation to System